\documentclass[prl, twocolumn, showpacs, showkeys, amsmath, amssymb]{revtex4}
\usepackage{graphicx}

\begin{document}

\title{Imperfect DNA Repair and the Error Catastrophe}

\author{Yisroel Brumer$^\dagger$}
\email{brumer@fas.harvard.edu}
\author{Emmanuel Tannenbaum\footnote{These authors contributed equally}}
\email{etannenb@fas.harvard.edu}
\author{Eugene I. Shakhnovich}
\affiliation{Harvard University, 12 Oxford Street, Cambridge, 
             Massachusetts 02138}

\begin{abstract}

In this Letter, we extend the semiconservative quasispecies equations to
incorporate imperfect DNA lesion repair.  We study the equilibrium behavior
of this model in the limit of infinite sequence length and population size,
using a single-fitness-peak landscape for which the master genome can sustain 
a finite number of lesions and remain viable.  We provide a full analytical
treatment of the problem, providing a general mathematical framework
as well as the full solution for a particular class of fitness landscapes. 
Stochastic simulations using finite sequence lengths and populations agree 
well with the analytical results.  Applications to biological systems are 
briefly discussed.

\end{abstract}

\pacs{87.14.Gg, 87.23.-n, 87.10.+e}

\keywords{Quasispecies, error catastrophe, lesion repair, semiconservative}

\maketitle
                                                                              
The quasispecies model of genomic evolution has been used to study a number
of problems in evolutionary dynamics.  The central result of the
theory is the existence of an upper mutational threshold beyond which
natural selection can no longer occur \cite{EIG1}.  Below this 
threshold, a replicating population of genomes will eventually produce, over 
many generations, a ``cloud'' of closely related genomes clustered about
one or a few fast replicating genomes.  These ``clouds'' are termed
quasispecies, and are characteristic of the evolutionary dynamics
of many viruses, such as HIV \cite{EIG2, LOEB, WILKE}. 

Above the mutational threshold, natural selection can no longer act to localize
the population about the fast replicating genomes, and delocalization occurs
over the entire genome space.  This localization to delocalization transition
is known as the error catastrophe \cite{EIG1, GALLUCCIO}, and it corresponds 
to the disappearance of any viable strains in the population.  The error 
catastrophe has been observed experimentally \cite{CAT1, CAT2}, and is 
believed to form the basis for a number of antiviral therapies 
\cite{EIG2, LOEB, CAT1}.

Because the quasispecies equations were originally developed to deal with
single-stranded RNA genomes, the model implicitly assumed a conservative
replication mechanism, where the original genome is preserved.  However, in 
order to apply the quasispecies model to living systems, whose genomes are 
DNA-based, it was necessary to develop the quasispecies equations for 
semiconservative replication.  In semiconservative replication, a 
double-stranded genome unzips to form two strands, each of which is used as a 
template for the formation of two new complementary strands by the rules of 
Watson-Crick base pairing \cite{VOET}.  The original genome is destroyed by 
this process, and because replication errors can happen in both daughter 
strand syntheses, it is possible that the two daughter genomes will
differ from the parent.

Daughter strand synthesis from the parent template strand is not error-free.
Therefore, living systems have evolved a host of mechanisms which correct 
base-pair mismatches during replication (some of these mechanisms are built 
into the DNA replicases themselves.  Others, such as mismatch repair, occur 
immediately following daughter strand synthesis) \cite{VOET}.  Nevertheless, 
after replication has occured, the daughter genomes may still contain 
mismatched base-pairs.  These mismatches result in lesions along the DNA chain,
which are repaired by DNA repair and maintenance enzymes present in the
cell.  Unlike repair that occurs during daughter strand synthesis, during 
lesion repair the parent and daughter strands are indistinguishable, and hence 
correct repair occurs with a probability of $ 1/2 $.

The semiconservative quasispecies equations were derived in \cite{SEMICONSERV},
under the simplifying asssumption that post-replication lesion repair is
perfectly efficient.  This Letter provides an extension of the original 
semiconservative quasispecies equations, to account for the case when lesion 
repair is imperfect (the full details of the solution presented in this work 
may be found in \cite{LESFULL}).  Such an extension is necessary for a proper 
modeling of many important biological processes.  Indeed, imperfect lesion 
repair was first studied in \cite{BRUMER3} in the context of modeling cancer.  
It may also be important for properly modeling assymetric stem cell kinetics 
(the so-called ``immortal strand'' hypothesis) \cite{SHERLEY}.

When lesion repair is perfectly efficient, double-stranded DNA consists of two 
complementary, antiparallel strands \cite{VOET, SEMICONSERV}.  Each DNA genome 
is defined by the pair of strands $ \{\sigma, \bar{\sigma}\} = \{\bar{\sigma}, 
\sigma\} $, where $ \bar{\sigma} $ denotes the complement of $ \sigma $.  If 
each base is drawn from an alphabet of size $ S $ (where $ S = 4 $ for
known terrestrial life), and if $ \bar{b}_i $ denotes the complement of a base 
$ b_i $, then if $ \sigma = b_1 \dots b_L $, we have, by the antiparallel 
nature of DNA, that $ \bar{\sigma} = \bar{b}_L \dots \bar{b}_1 $. 

The replication of a DNA genome $ \{\sigma, \bar{\sigma}\} $ may be
divided into three stages:  (1)  Strand separation, where the genome
unzips to produce two parent strands, $ \sigma $ and $ \bar{\sigma} $.
(2)  Daughter strand synthesis, where each parent strand serves as the
template for the synthesis of a complementary daughter strand.  (3)  Lesion
repair after cell division.  This replication mechanism leads to the
semiconservative quasispecies equations developed in \cite{SEMICONSERV}.

When lesion repair is imperfect, the correlation between the two strands is 
broken, and we must consider a more generalized dynamics over genomes of the 
form $ \{\sigma, \sigma'\} $, where both $ \sigma $ and $ \sigma' $ are 
arbitrary.  Following the derivation in \cite{SEMICONSERV}, we obtain
the quasispecies equations
\begin{eqnarray}
\frac{d x_{\{\sigma, \sigma'\}}}{dt} 
& = &
-(\kappa_{\{\sigma, \sigma'\}} + \bar{\kappa}(t)) 
x_{\{\sigma, \sigma'\}} \nonumber \\
&   &
+ \sum_{\{\sigma'', \sigma'''\}}
\kappa_{\{\sigma'', \sigma'''\}} x_{\{\sigma'', \sigma'''\}} \times 
\nonumber \\ 
&   &
[p((\sigma'', \sigma'''), \{\sigma, \sigma'\}\}) + 
 p((\sigma''', \sigma''), \{\sigma, \sigma'\}\})] \nonumber \\
\end{eqnarray}
where $ p((\sigma'', \sigma'''), \{\sigma, \sigma'\}) $ denotes the probability
that strand $ \sigma'' $, as part of genome $ \{\sigma'', \sigma'''\} $, 
becomes genome $ \{\sigma, \sigma'\} $ after daughter strand synthesis and 
lesion repair.  Here $ x_{\{\sigma, \sigma'\}} $ denotes the fraction of the 
population with genome $ \{\sigma, \sigma'\} $, and $ \bar{\kappa}(t) \equiv
\sum_{\{\sigma, \sigma'\}}{\kappa_{\{\sigma, \sigma'\}} 
x_{\{\sigma, \sigma'\}}} $ is the mean fitness of the population.

In the semiconservative quasispecies equations, the complementarity 
property allows one to convert the quasispecies dynamics over 
double-stranded genomes into an equivalent (and considerably simpler)
dynamics over single strands \cite{SEMICONSERV}.  With imperfect
lesion repair, the lack of perfect correlation between the
two strands in the genome makes a conversion to a single strand
model impossible.  Nevertheless, we can make an analogous 
transformation of the dynamics, from double-stranded genomes
$ \{\sigma, \sigma'\} $ to {\it orderered pairs} of strands,
$ (\sigma, \sigma') $, as follows:  We define $ y_{(\sigma, \sigma')} = 
y_{(\sigma', \sigma)} = \frac{1}{2} x_{\{\sigma, \sigma'\}} $ if 
$ \sigma \neq \sigma' $, and $ y_{(\sigma, \sigma)} = x_{\{\sigma, \sigma\}} $.
Also, we define $ \kappa_{(\sigma, \sigma')} = \kappa_{(\sigma', \sigma)} =
\kappa_{\{\sigma, \sigma'\}} $.  Finally, we define $ p((\sigma'', \sigma'''),
(\sigma, \sigma')) $ to be the probability that $ \sigma'' $, as part
of genome $ \{\sigma'', \sigma'''\} $, becomes $ \sigma $, with daughter 
strand $ \sigma' $ (after daughter strand synthesis and lesion repair).
Then it follows that
\begin{equation}
p((\sigma'', \sigma'''), \{\sigma, \sigma'\}) =
\left\{ \begin{array}{cc}
        p((\sigma'', \sigma'''), (\sigma, \sigma')) + \\
        p((\sigma'', \sigma'''), (\sigma', \sigma)) &
	\mbox{if $ \sigma \neq \sigma' $} \\
	p((\sigma'', \sigma'''), (\sigma, \sigma')) &
	\mbox{if $ \sigma = \sigma' $}
	\end{array}
\right.				  
\end{equation}
Using these definitions, it is possible to convert the quasispecies equations
over the space of double-stranded genomes to the space of ordered sequence
pairs.  After some manipulation, the final result is,
\begin{eqnarray}
\frac{d y_{(\sigma, \sigma')}}{dt} 
& = &
-(\kappa_{(\sigma, \sigma')} + \bar{\kappa}(t)) y_{(\sigma, \sigma')}
\nonumber \\
&   &
+ \sum_{(\sigma'', \sigma''')}
\kappa_{(\sigma'', \sigma''')} y_{(\sigma'', \sigma''')} \times \nonumber \\
&   &
[p((\sigma'', \sigma'''), (\sigma, \sigma')) + 
 p((\sigma'', \sigma'''), (\sigma', \sigma))]. \nonumber \\
\end{eqnarray}

To determine $ p((\sigma'', \sigma'''), (\sigma, \sigma')) $, we introduce 
some additional definitions.  Define $ \sigma_C $ to be the subsequence of 
bases in $ \sigma $ which are complementary with the corresponding bases in 
$ \sigma' $.  That is, suppose $ \sigma = b_1 \dots b_L $, and suppose for 
indices $ i_1 < i_2 < \dots i_k $ we have that $ \bar{b}_{i_j} = 
b_{L - i_j + 1}' $.  Then $ \sigma_C = b_{i_1} \dots b_{i_k} $.  We also 
define $ \sigma_C' $ to be the subsequence of corresponding bases in 
$ \sigma' $, so that $ \sigma_C' = b_{L-i_k+1}' \dots b_{L-i_1+1}' $.  
Finally, let $ \sigma_C'' $ denote the subsequence of bases in $ \sigma'' $ 
corresponding to the bases in $ \sigma_C $, so that $ \sigma_C'' = b_{i_1}'' 
\dots b_{i_k}'' $.

Now, define $ \sigma_{NC} $ to be the subsequence of bases in $ \sigma $ which
are not complementary with the corresponding bases in $ \sigma' $.  That is,
given the complementary indices $ i_1 < i_2 < \dots < i_k $ defined above,
let $ i_1' < i_2' < \dots < i_{L-k}' $ be the remaining indices.  Then
$ \sigma_{NC} = b_{i_1'} \dots b_{i_{L-k}'} $.  We define $ \sigma_{NC}' $
to be the subsequence of corresponding bases in $ \sigma' $, so that
$ \sigma_{NC}' = b_{L-i_{L-k}'+1} \dots b_{L-i_1'+1} $.  Finally,
we let $ \sigma_{NC}'' $ denote the subsequence of bases in $ \sigma'' $ 
corresponding to the bases in $ \sigma_{NC} $, so that $ \sigma_{NC}'' =
b_{i_1'} \dots b_{i_{L-k}'} $.

We also assume that daughter strand synthesis during replication of the
genome $ \{\sigma'', \sigma'''\} $ is characterized by a per base-pair
mismatch probability of $ \epsilon_{\{\sigma'', \sigma'''\}} $, and we
define $ \epsilon_{(\sigma'', \sigma''')} = \epsilon_{(\sigma''', \sigma'')} =
\epsilon_{\{\sigma'', \sigma'''\}} $.  Finally, we define $ \lambda $
to be the probability that a post-replicative lesion is repaired. 
This gives,
\begin{widetext}
\begin{equation}
p((\sigma'', \sigma'''), (\sigma, \sigma')) = 
\delta_{\sigma''_{NC} \sigma_{NC}}
(\frac{\lambda \epsilon_{(\sigma'', \sigma''')}}{2 (S - 1)})^
{D_H(\sigma_C'', \sigma_C)}  
(1 - \epsilon_{(\sigma'', \sigma''')} (1 - \frac{\lambda}{2}))^
{L - D_H(\sigma, \bar{\sigma}') - D_H(\sigma_C'', \sigma_C)} 
(\frac{\epsilon_{(\sigma'', \sigma''')} (1 - \lambda)}{S-1})^
{D_H(\sigma, \bar{\sigma}')}.
\end{equation}
\end{widetext}
For $ \lambda = 1 $ (all lesions are repaired), our equations reduce to the 
ordinary semiconservative quasispecies equations \cite{SEMICONSERV}.

In the simplest case, we assume that $ \epsilon_{\{\sigma, \sigma'\}} $
is genome-independent, and hence may be denoted by $ \epsilon $.  We also
define $ \mu = L\epsilon $, and consider the quasispecies dynamics at fixed
$ \mu $ in the limit of $ L \rightarrow \infty $.  Note that 
$ \lim_{L \rightarrow \infty, L\epsilon = \mu}{(1 - \epsilon)^L} = e^{-\mu} $, 
so fixing $ \mu $ is equivalent to holding the correct daughter strand 
synthesis probability constant in the limit of infinite sequence length. 

We now consider a generalized ``single-fitness peak'' landscape, characterized
by a ``master'' genome $ \{\sigma_0, \bar{\sigma}_0\} $.  A given genome
$ \{\sigma, \sigma'\} $ is viable, with a first-order growth rate constant
$ k > 1 $, if it is equal to the master genome, differing by at most $ l $
lesions.  Otherwise, the genome is unviable, with a growth rate constant of 
$ 1 $.

In the limit of infinite sequence length, it may be shown that, with 
probability one, the Hamming distance between $ \sigma_0 $ and 
$ \bar{\sigma}_0 $ is infinite \cite{SEMICONSERV}.  Therefore, we may regard 
$ (\sigma_0, \bar{\sigma}_0) $ and $ (\bar{\sigma}_0, \sigma_0) $ as 
infinitely separated in the sequence-pair space, and so, by an appropriate 
transformation of Eq. (3), we may consider the local dynamics 
about each sequence pair independently of the other.  Thus, we consider 
the dynamics of the $ x_{(\sigma, \sigma')} $ for two types of 
$ (\sigma, \sigma') $:  First, we consider $ (\sigma, \sigma') $ such that 
$ D_H(\sigma, \sigma_0) $, $ D_H(\sigma', \bar{\sigma}_0) $ are finite, and 
second, we consider $ (\sigma, \sigma') $ such that 
$ D_H(\sigma, \bar{\sigma}_0) $, $ D_H(\sigma', \sigma_0) $ are finite.  If 
$ (\sigma, \sigma') $ belongs to the first type of sequence pairs, then it is 
clear that $ (\sigma', \sigma) $ belongs to the second type.  The symmetry of 
the landscape means that the dynamics about one sequence pair completely 
determines the dynamics about the other.

A given sequence pair $ (\sigma, \sigma') $ of the first type can be
characterized by the four parameters $ l_C $, $ l_L $, $ l_R $, and $ l_B $.
The first parameter, $ l_C $, denotes the number of positions where
$ \sigma $, $ \sigma' $ are complementary, yet differ from the corresponding
positions in $ \sigma_0 $, $ \bar{\sigma}_0 $, respectively.  The second
parameter, $ l_L $, denotes the number of positions where $ \sigma $ differs 
from $ \sigma_0 $, but the complementary positions in $ \sigma' $ are equal 
to the corresponding ones in $ \bar{\sigma}_0 $.  The third parameter,
$ l_R $, denotes the number of positions where $ \sigma $ is equal to the
ones in $ \sigma_0 $, but the complementary positions in $ \sigma' $ differ
from the corresponding ones in $ \bar{\sigma}_0 $.  Finally, the fourth 
parameter, $ l_B $, denotes the number of positions where $ \sigma $, 
$ \sigma' $ are not complementary, and also differ from the corresponding 
positions in $ \sigma_0 $ and $ \bar{\sigma}_0 $, respectively.

For our generalized single-fitness peak model, the fitness of a given sequence 
pair $ (\sigma, \sigma') $ of the first type is determined by $ l_C $, 
$ l_L $, $ l_R $, $ l_B $, hence we may write $ \kappa_{(\sigma, \sigma')} = 
\kappa_{(l_C, l_L, l_R, l_B)} $.  Specifically, 
$ \kappa_{(l_C, l_L, l_R, l_B)} = k > 1 $ if $ l_C = 0 $, and if 
$ l_L + l_R + l_B \leq l $.  Otherwise, $ \kappa_{(l_C, l_L, l_R, l_B)} = 1 $.

We define $ z_{(l_C, l_L, l_R, l_B)} $ to be the total fraction of the 
population whose genomes are characterized by the parameters $ l_C $, $ l_L $, 
$ l_R $, $ l_B $.  Note that we can consider these same parameters as 
characterizing genomes of the second type (i.e., defined by the ordered pair 
$ (\bar{\sigma}_0, \sigma_0) $), and consider the corresponding population
fraction $ \bar{z}_{(l_C, l_L, l_R, l_B)} $.  It should be clear, that,
by symmetry, $ \bar{z}_{(l_C, l_L, l_R, l_B)} = z_{(l_C, l_R, l_L, l_B)} $.

Because the fitness is only determined by $ l_C $, $ l_L $, $ l_R $,
and $ l_B $, it follows that we may presymmetrize our population and
reexpress the quasispecies dynamics in terms of the $ z_{(l_C, l_L, l_R,
l_B)} $.  In \cite{LESFULL}, we show that the neglect of backmutations
in the limit of infinite sequence length implies that we may set $ z_{(l_C,
l_L, l_R, l_B)} = 0 $ when $ l_B \neq 0 $, and when $ l_L, l_R $ are 
simultaneously nonzero.  Therefore, the relevant equations are
\begin{widetext}
\begin{eqnarray}
\frac{d z_{(l_C, 0, 0, 0)}}{dt} 
& = &
-(\kappa_{(l_C, 0, 0, 0)} + \bar{\kappa}(t)) z_{(l_C, 0, 0, 0)} \nonumber \\
&   &
+ 2 e^{-\mu (1 - \lambda/2)} 
\sum_{l_C' = 0}^{l_C} \frac{1}{l_C'!} (\frac{\lambda \mu}{2})^{l_C'}
\sum_{l_1'' = 0}^{l_C - l_C'} 
\sum_{l_2'' = 0}^{\infty}
\kappa_{(l_1'', l_C - l_C' - l_1'', l_2'', 0)} 
z_{(l_1'', l_C - l_C' - l_1'', l_2'', 0)} \\
\frac{d z_{(l_C, l_L, 0, 0)}}{dt} 
& = &
-(\kappa_{(l_C, l_L, 0, 0)} + \bar{\kappa}(t)) z_{(l_C, l_L, 0, 0)} 
\nonumber \\
&   &
+ \frac{1}{l_L!} (\mu(1 - \lambda))^{l_L} e^{-\mu (1 - \lambda/2)}
\sum_{l_C' = 0}^{l_C} \frac{1}{l_C'!} (\frac{\lambda \mu}{2})^{l_C'}
\sum_{l_1'' = 0}^{l_C - l_C'} 
\sum_{l_2'' = 0}^{\infty}
\kappa_{(l_1'', l_C - l_C' - l_1'', l_2'', 0)} 
z_{(l_1'', l_C - l_C' - l_1'', l_2'', 0)} \\
\frac{d z_{(l_C, 0, l_R, 0)}}{dt} 
& = &
-(\kappa_{(l_C, 0, l_R, 0)} + \bar{\kappa}(t)) z_{(l_C, 0, l_R, 0)} 
\nonumber \\
&   &
+ \frac{1}{l_R!} (\mu(1 - \lambda))^{l_R} e^{-\mu (1 - \lambda/2)}
\sum_{l_C' = 0}^{l_C} \frac{1}{l_C'!} (\frac{\lambda \mu}{2})^{l_C'}
\sum_{l_1'' = 0}^{l_C - l_C'} 
\sum_{l_2'' = 0}^{\infty}
\kappa_{(l_1'', l_C - l_C' - l_1'', l_2'', 0)} 
z_{(l_1'', l_C - l_C' - l_1'', l_2'', 0)} \nonumber \\
\end{eqnarray}
\end{widetext}

The above equations may be used to solve for the equilibrium mean
fitness $ \bar{\kappa}(t = \infty) $ for the generalized single-fitness-peak 
landscape.  Below the error catastrophe, the result is
\begin{equation}
\bar{\kappa}(t = \infty) = \frac{A(\mu, \lambda) + 
\sqrt{A(\mu, \lambda)^2 + 4 B(\mu, \lambda)}}{2}
\end{equation}
where $ A(\mu, \lambda) = k((1 + f_l(\mu, \lambda)) e^{-\mu (1 - \lambda/2)} 
- 1) - f_l(\mu, \lambda) e^{-\mu (1 - \lambda/2)} + e^{-\mu \lambda/2} - 1 $,
and $ B(\mu, \lambda) = k(e^{-\mu \lambda/2} + e^{-\mu (1 - \lambda/2)} - 1) $,
where we define $ f_l(\mu, \lambda) = \sum_{l' = 0}^{l}{\frac{1}{l'!}
[\mu (1 - \lambda)]^{l'}} $. 

The error catastrophe occurs when the mean equilibrium fitness determined
by Eq. (8) becomes equal to the growth rate of the unviable genomes.  At
this point, the selective advantage of the viable genomes is no longer 
sufficiently strong to localize the population, and delocalization occurs
over the entire genome space.  The critical $ \mu $ is therefore found
by setting $ \bar{\kappa}(t = \infty) = 1 $ in Eq. (8), and solving for
$ \mu $.  The resulting expression is
\begin{equation}
\frac{e^{-\mu (1 - \lambda/2)}}{2 - e^{-\mu \lambda/2}} =
\frac{k+1}{k(2 + f_l(\mu, \lambda)) - f_l(\mu, \lambda)}.
\end{equation}

It is instructive to study the behavior of $ \bar{\kappa}(t = \infty) $ for 
specific landscapes and values of $ \lambda $.  First of all, note that 
$ f_l(\mu, 1) = 1 $, giving $ \bar{\kappa}(t = \infty) = k(2 e^{-\mu/2} - 1) 
$, which is exactly the expected semiconservative result with perfect lesion 
repair \cite{SEMICONSERV}.  Also, note that $ f_{\infty}(\mu, \lambda) = 
e^{(1 - \lambda) \mu} $, which gives $ \bar{\kappa}(t = \infty) = 
k(e^{-\mu(1 - \lambda/2)} + e^{-\mu \lambda/2} - 1) $.  For $ \lambda = 1 $,
we of course recover the semiconservative result.  However, for 
$ \lambda = 0 $, we obtain $ \bar{\kappa}(t = \infty) = k e^{-\mu} $,
which is exactly the result expected from conservative replication.  Therefore,
when only one perfect strand in a double stranded genome is necessary for the 
organism to remain viable, we recover an effectively conservatively 
replicating system in the absence of lesion repair (see also \cite{BRUMER3}).  

In Figure 1 we show some results of stochastic simulations of replicating
genomes, which corroborate the analytical results obtained from our theory.

\begin{figure}
\includegraphics[width = 0.85\linewidth, angle = -90]{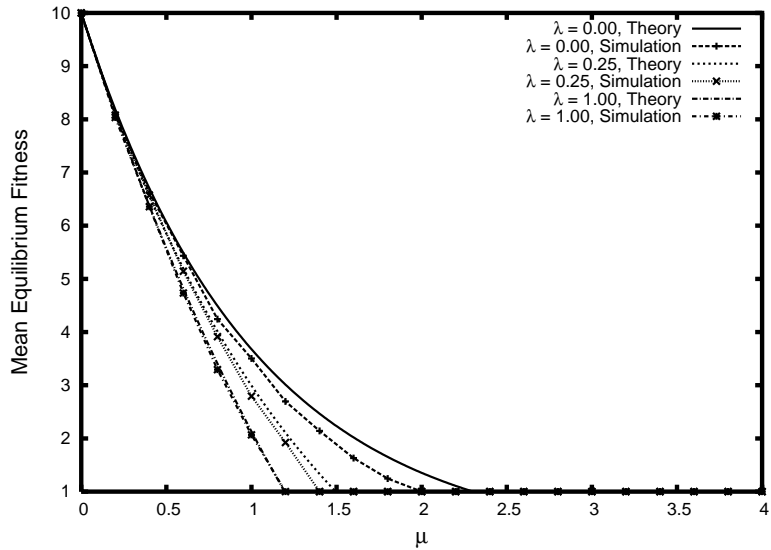}
\caption{Plots of $ \bar{\kappa}(t = \infty) $ versus $ \mu $, from both
stochastic simulation and theory.  We took $ l = \infty $.  For our 
stochastic simulations, we averaged our results over $ 10 $ runs, using
sequence lengths of $ 20 $, and a population size of $ 1,000 $ organisms.}
\end{figure}

The recent incorporation of semiconservative replication into the quasispecies 
model was an important step toward modeling real systems that revealed a 
number of important dynamical signatures absent in the original model.
However, the initial assumption used in previous semiconservative works, 
namely that post-replication DNA repair is perfect, is clearly an 
oversimplification that is particularly poor for some of the most
scientifically interesting systems such as cancer and stem cells
\cite{BRUMER3, LESFULL, SHERLEY}. This approximation introduces a false 
symmetry that can drastically alter the evolutionary behavior and equilibria.  
By providing a full treatment of semiconservative quasispecies dynamics with 
partially activated lesion repair, we have taken a significant step forward in 
the modeling of genomic evolution.

\begin{acknowledgments}

This research was supported by the National Institutes of Health.  The authors
would like to thank Prof. James L. Sherley and Franziska Michor for useful
discussions.

\end{acknowledgments}

\end{document}